\documentclass[aps,prb,showpacs,twocolumn]{revtex4}
\usepackage{graphicx}
\usepackage{bm}
\begin{document}
\title{Nonlinear magnetoresistance  
of an irradiated two-dimensional electron system}
\author{X. L. Lei and S. Y. Liu}
\affiliation{Department of Physics, Shanghai Jiaotong University,
1954 Huashan Road, Shanghai 200030, China}

\begin{abstract}
Nonlinear magnetotransport of a microwave-irradiated high mobility two-dimensional 
electron system under a finite direct current excitation 
is analyzed using a dc-controlled scheme with photon-assisted transition mechanism.
The predicted amplitudes, extrema and nodes of the oscillatory 
differential resistance versus the magnetic field and the current density,  
are in excellent agreement with the recent 
experimental observation [Hatke {\it et al.} Phys. Rev. B {\bf 77}, 201304(R) (2008)].

\end{abstract}


\maketitle

The prediction\cite{Ryz-1970} and detection\cite{Zud01,Ye} of radiation 
induced magnetoresistance oscillation (RIMO) in two-dimensional (2D) electron 
systems (ES), especially the discovery of the zero-resistance state\cite{Mani02,Zud03},  
have stimulated intensive
experimental\cite{Dor03,Yang03,Zud04,Mani04,Willett,Du,Kovalev,Mani05,Dor05,
Stud,Smet05,Yang06,Bykov05,Bykov06} 
and theoretical\cite{Shi,Durst,Lei03,Lei04,Ryz03,Vav04,Dmitriev03,DGHO05,Torres05,
Ng05,Ina-prl05,Kashuba,Andreev03,Alicea05,Auerbach05,Mikhailov04} studies on  
this extraordinary transport phenomenon of electrons in very high Landau levels. 

Despite the fact that basic features of RIMO have been established 
and the understanding that it stems  
from impurity scattering has been reached, 
so far there has been no common agreement as to the accurate microscopic origin 
of these giant resistance oscillations. 
Presented in different forms, many theoretical models\cite{Shi,Durst,Lei03,Lei04,Ryz03,Vav04,Torres05,
Ina-prl05,Kashuba,Ng05,Lei07-2,Auerbach07} consider
RIMO to arise from electron transitions between different Landau states
due to impurity scattering accompanied by absorbing and emitting microwave photons. 
This origin is called the "photon-assisted transition" or "displacement" mechanism. 
A different origin, called the "inelastic" or "distribution function" mechanism,\cite{Dor03,Dmitriev03,DGHO05} 
considers RIMO to arise from a microwave-induced nonequilibrium oscillation 
of the time-averaged isotropic electron distribution function 
in the density-of-states (DOS) modulated system.
Both mechanisms exist in a real 2D semiconductor and have been shown to 
produce magentoresistance oscillations qualitatively having the observed
period, phase and magnetic field damping. The "displacement" mechanism predicts a well-defined 
photoresistivity with given impurity scattering and Landau-level broadening, while
the "inelastic" mechanism yields an additional factor   
proportional to the ratio of the inelastic scattering time $\tau_{\rm in}$ 
to the impurity-induced quantum scattering time $\tau_{q}$.\cite{Dmitriev03} 
Since the inelastic scattering time $\tau_{\rm in}$ or the thermalization time $\tau_{\rm th}$,\cite{Lei03}
being the property of a nonequilibrium state and contributed 
by the direct Coulomb interactions between electrons and by all other
possible impurity- and phonon-scattering mediated effective
electron-electron scatterings,\cite{Lei03} is very hard 
to determine theoretically or to measure experimentally,
 the sharp controversy whether $\tau_{\rm in}/\tau_{q}\gg 1$ or 
$\tau_{\rm in}/\tau_{q}\ll 1$, i.e. which mechanism plays the dominant role 
in the experimental systems,\cite{Lei03,Dmitriev03} has been an unsolved issue.
The detailed comparison between theoretical predictions and experiments 
may provide a useful way to distinguish them.

Introducing additional parameters into microwave-illuminated 2DESs, such as dc excitations,
can be of help to distinguish different models and mechanisms. 
It has been shown that a finite current alone, can also induce substantial magnetoresistance oscillation
and zeroresistance without microwave radiation.\cite{Yang02,WZhang07,JZhang07,Bykov07,Lei07-1,Vav07}
Simultaneous application of a direct current and a microwave radiation leads to very  
interesting and complicated oscillatory behavior of resistance and 
differential resistance.\cite{WZhang07-2,Lei07-2,Auerbach07}
Recent careful measurements\cite{WZhang08,Hatke08}
disclosed further details of such nonlinear magnetotransport in a high-mobility 2D semiconductor 
under both ac and dc exitations, allowing a careful comparison with theoretical predictions.  
 
Our examination is based on a current-controlled scheme of photon-assisted transport,\cite{Lei03}
which deals with a 2DES of short thermalization time having $N_{\rm s}$ electrons in a unit area of 
the $x$-$y$ plane and subject to a uniform magnetic field ${\bm B}=(0,0,B)$ 
in the $z$ direction.
When an electromagnetic wave with incident electric field ${\bm E}_{{\rm i}s}\sin \omega t$
irradiates perpendicularly on the plane together with a dc electric field ${\bm E}_0$ inside,
the steady transport state of this 2DES is described by the electron drift velocity
${\bm v}_0$ and an electron temperature $T_{\rm e}$, satisfying the 
force and energy balance equations\cite{Lei03}
\begin{eqnarray}
N_{\rm s}e{\bm E}_{0}+N_{\rm s} e ({\bm v}_0 \times {\bm B})+
{\bm F}_0&=&0,\label{eqv0}\\
N_{\rm s}e{\bm E}_0\cdot {\bm v}_0+S_{\rm p}- W&=&0.
\label{eqsw}
\end{eqnarray}
Here, the frictional force  resisting electron drift motion,
\begin{equation}
{\bm F}_0=\sum_{{\bm q}_\|}\left| U({\bm q}_\|)\right| ^{2}
\sum_{n=-\infty }^{\infty }{\bm q}_\|{J}_{n}^{2}(\xi ){\Pi}_{2}
({\bm q}_\|,\omega_0+n\omega ),\label{ff0}
 \label{eqf0}
\end{equation}
is given in terms of the electron density correlation function ${\Pi}_2({\bm q}_{\|},{\Omega})$,
the effective impurity potential $U({\bm q}_{\|})$, a radiation-related coupling parameter
$\xi$ in the Bessel function $J_n(\xi)$, and $\omega_0\equiv{\bm q}_{\|}\cdot {\bm v}_0$.
The electron energy absorption from the radiation field, $S_{\rm p}$, and the electron energy dissipation 
to the lattice, $W$, are given in Ref.\,\onlinecite{Lei03}.
The nonlinear longitudinal resistivity and differential resistivity 
in the presence of a radiation field are obtained from Eq.\,(\ref{eqv0}) 
by taking ${\bm v}_0$ and the current density ${\bm J}=N_{\rm s}e{\bm v}_0 $ 
in the $x$ direction, ${\bm v}_0=(v_{0},0,0)$ and ${\bm J}=(J,0,0)$,
\begin{equation}
R_{xx}=-{F}_0/(N_{\rm s}^2 e^2 v_{0}),\,\,\,
r_{xx}=-({\partial F_0}/{\partial v_0})/(N_{\rm s}^2 e^2).\label{rrxx}
\end{equation}

We have calculated the differential resistivity $r_{xx}$ from above equations 
(taking up to three-photon processes) under different magnetic fields $B$ 
and bias drift velocities $v_0$ for a GaAs-based heterosystem  
with carrier density $N_{\rm s}=3.7\times 10^{15}$/m$^{2}$ and low-temperature
linear mobility $\mu_0=1200$\,m$^{2}$/V\,s at lattice temperature $T=1.5$\,K, 
irradiated by a linearly $x$-polarized microwave of frequency $\omega/2\pi=69$\,GHz 
with incident amplitude $E_{{\rm i}s}=3.6$\,V/cm. 
The elastic scatterings are assumed due to a mixture
of short-range and background impurities, and the Landau-level broadening $\Gamma$ is taken 
to be a Gaussian form with a broadening parameter $\alpha=7$.\cite{Lei03}

\begin{figure}
\includegraphics [width=0.45\textwidth,clip=on] {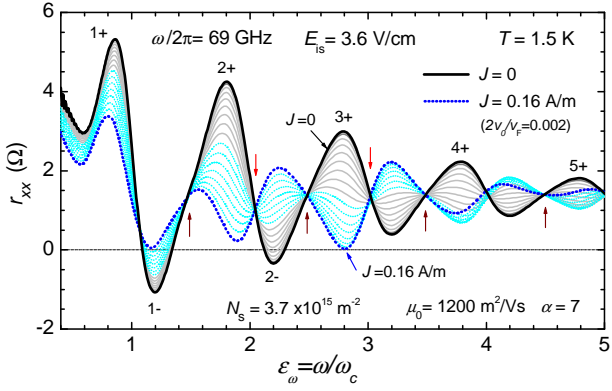}
\vspace*{-0.2cm}
\caption{(Color online) Differential magnetoresistivity $r_{xx}$ vs $\epsilon_{\omega}=\omega/\omega_c$
under fixed bias current densities from $J=0$ to $J=0.16$ A/m, in $0.01$\,A/m increments.}
\label{fig1}
\end{figure}
\begin{figure}
\includegraphics [width=0.45\textwidth,clip=on] {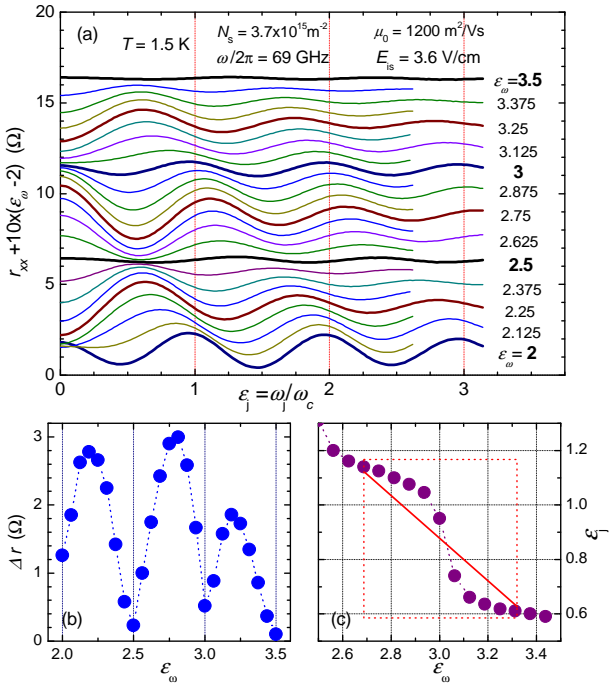}
\vspace*{-0.2cm}
\caption{(Color online) (a) Differential resistivity $r_{xx}$ vs $\epsilon_j$
at fixed $\epsilon_{\omega}$ from 2 to 3.5 in steps of 0.0625.
(b) Amplitude of $r_{xx}$ oscillations. (b) Positions of $r_{xx}$ maxima extracted.}
\label{fig2}
\end{figure}

Figure 1 presents 
the calculated $r_{xx}$ versus $\epsilon_{\omega}\equiv \omega/\omega_{c}$
($\omega_c=eB/m$ is the cyclotron frequency) 
at fixed bias drift velocities  from 
$2v_0/v_{\rm F}=0$ ($v_{\rm F}$ is the Fermi velocity) to $2\times 10^{-3}$ in steps of $1.25\times 10^{-4}$,
corresponding to current densities $J=0$ to $0.16$\,A/m in steps of $0.01$\,A/m.
The $J=0$ case exhibits typical RIMO with a sequence of resistance maxima ($1^+,2^+,3^+,4^+$ and $5^+$)
and negative values around the resistance minima $1^-$ and $2^-$. 
With increasing $J$ to 0.16\,A/m, the maxima $2^+,3^+$ and $4^+$ (minima $2^-,3^-$ and $4^-$) 
evolve into minima (maxima) having seemingly little change in the $B$ positions. 
Further, all the curves cross approximately at $\epsilon_{\omega}=1.5,2,2.5,3,3.5$ and 4.5,
indicating that $J$ at this range has little effect on photoresistance there. 
These and other features of Fig.\,1 reproduce what was exactly observed in Ref.\onlinecite{Hatke08}.    

Figure 2(a) shows the calculated $r_{xx}$ versus  
 $\epsilon_j \equiv \omega_j/\omega_c$ ($\omega_j=2k_{\rm F}v_0$, $k_{\rm F}$ is the Fermi wave vector
of the 2D electron system) at fixed $\epsilon_{\omega}$ 
from $2$ to $3.5$ in steps of 0.0625. Traces are vertically offset in increments of $0.625\,\Omega$ 
for clarity. The amplitudes $\Delta r$ of $r_{xx}$-$\epsilon_j$ oscillations are 
maximized around $\epsilon_{\omega}\simeq 2.2,2.8$ and 3.2 and strongly suppressed at 
$\epsilon_{\omega} \simeq 2.5$ and 3.5, as shown in Fig.\,2(b).\cite{note} 
The positions of $r_{xx}$ maxima extracted are plotted in Fig.\,2(c) as 
dots. All these are in excellent agreement with the experimental results
[Fig.\,2(a),(b) and (c) of Ref.\onlinecite{Hatke08}].

In the present current-controlled transport model
the oscillations of $R_{xx}$ and $r_{xx}$ are referred to the behavior of
function ${\Pi}_{2}({\bm q}_\|,\omega_0+n\omega )$ in Eq.\,(\ref{ff0}).
The electron density correlation function ${\Pi}_{2}({\bm q}_\|,\Omega )$ 
is essentially a multiplication 
of two energy-$\Omega$ shifted periodically modulated DOS functions 
of electrons in the magnetic field.\cite{Lei03} 
Its periodicity with changing frequency $\Omega\rightarrow \Omega+\omega_c$ at low temperatures 
and high Landau-level occupations, determines the main periodical behavior of magnetoresistance.\cite{Lei03} 
The previous examination\cite{Lei07-2} focused on the node positions of the oscillatory peak-valley
pairs of $R_{xx}$, which appear periodically
roughly along the lines
$ 
\epsilon_\omega+\eta\, \epsilon_j=m=1,2,3,4,... 
$
in the $\epsilon_\omega$-$\epsilon_j$ plane,  where $\epsilon_\omega\equiv\omega/\omega_c$
 is the control parameter of RIMOs,
and $\epsilon_j \equiv \omega_j/\omega_c$  
is the control parameter of current-induced magnetoresistance oscillations, 
and $\eta \lesssim 1$, dependent on the 
scattering potential.\cite{Lei07-1}

The maxima of differential resistivity $r_{xx}$ show up at lower values in the $\epsilon_j$ axis
in comparison with the node positions of related valley-peak pairs of $R_{xx}$,
and its appearance exhibits a periodicity $\Delta\epsilon_j \lesssim 1$.\cite{Lei07-1} 
In the $\epsilon_\omega$-$\epsilon_j$ plane, 
the differential resistance maxima are expected to show up roughly in the vicinity along the lines
\begin{equation} 
\epsilon_\omega+\lambda\, \epsilon_j=m=1,2,3,4,... \label{lambdam}
\end{equation}  
with $\lambda \gtrsim 1$, dependent on the scattering potential 
($\lambda = 1.04$ for the system on discussion).
Eq.\,(\ref{lambdam}) qualitatively accounts for the periodical change of $r_{xx}$ 
in a large scale in steps of $\Delta(\epsilon_\omega+\lambda\, \epsilon_j)=1$. 
  
Under strong microwave irradiation, as in the present case, 
the role of virtual photon process [the $n=0$ term
in the sum of Eq.\,(\ref{ff0})] is negligible due to samll $J_0^{2}(\xi)$,\cite{Lei03} 
and main contributions to resistivity come from $n=\pm1,\pm2,...$ terms
(single- and multiple-photon processes). Noticing that the frequency differentiate 
${\Pi}_{2}^{\prime}({\bm q}_\|,\Omega)$ is an even function of $\Omega$ and
considering contributions from scatterings 
parallel and antiparallel to the drift velocity ${\bm v}_0$ 
and from $\pm |n|$ terms, we see that, in the case of finite bias current, the $r_{xx}$ behavior is
determined by the sum of two terms: (a) ${\Pi}_{2}^{\prime}({\bm q}_\|,|n|\omega+q_{\|}v_0\cos{\theta})$
and (b) ${\Pi}_{2}^{\prime}({\bm q}_\|,|n|\omega-q_{\|}v_0\cos{\theta})$.
Depending on the  ${\Pi}_{2}^{\prime}({\bm q}_\|,\Omega)$ function behavior in the vicinity of
$\Omega=|n|\omega$, effects of these two terms 
can be cancelled or added, completely or partly, at different locations of $\epsilon_{\omega}$.
${\Pi}_{2}^{\prime}({\bm q}_\|,\Omega )$ function 
reaches maxima (positive) at around $\Omega/\omega_c=N-\frac{1}{4}$, reaches minima (negative) at around
 $\Omega/\omega_c=N+\frac{1}{4}$, and  
passes through zero (changing sign) at around $\Omega/\omega_c=N$ and $N+\frac{1}{2}$ for all
integers $N\geq 2$. Thus, at $\epsilon_{\omega}\simeq l$ or $l+\frac{1}{2}$ ($l=2,3,4,...$) with which all
involved $|n|\omega$ frequencies are located around $N\omega_c$ or $(N+\frac{1}{2})\omega_c$, 
contributions from (a) and (b) are almost cancelled out for modest $v_0$.
In the case of $\epsilon_{\omega}\simeq l-\frac{1}{4}$\, [$\epsilon_{\omega}\simeq l+\frac{1}{4}$], 
there always exists a term of frequency
$(N-\frac{1}{4})\omega_c$\, [($N+\frac{1}{4})\omega_c$] in $|n|\omega$, 
and contributions from (a) and (b) are positively [negatively]
additive. These clearly account for the suppression 
of the current effect at $\epsilon_{\omega}\simeq l$ and $l+\frac{1}{2}$,  
and  the enhancement of it around $\epsilon_{\omega} \simeq l-\frac{1}{4}$
and $\epsilon_{\omega}\simeq l+\frac{1}{4}$.

Above discussions are general. 
The accurate behavior of resistivity $r_{xx}$ inside a period scale is relevant to the detailed shape of 
the DOS function. Figs.\,1 and 2 represent the result of a Gaussian-type DOS. 
The good quantitative agreement with experiment without adjusting parameters 
indicates that the present current-controlled scheme of photon-assisted transport captures the main physics of RIMOs 
in the discussed quasi-2D system. 

This work was supported by the projects of the National Science Foundation of China,
and the Shanghai Municipal Commission of Science and Technology.


\begin{thebibliography}{99}
\bibitem{note} The amplitude maxima will be at $\epsilon_{\omega}\simeq 2.25,2.75$ and 3.25 
if one considers only up to single-photon process ($|n|\leq 1$).
\end{thebibliography}
\end{document}